\documentstyle[12pt]{article}

\newcommand\mathC{{\mkern1mu\raise2.2pt\hbox{$\scriptscriptstyle|$}
        {\mkern-7mu\rm C}}}
\newcommand{\mathR}{{\rm I\! R}}                

\renewcommand\[{[\,}                            
\renewcommand\]{\,]}                            

\renewcommand\mathR{{\rm I\! R}}

\newcommand\unit{{\rm 1\kern-3.2pt I}}

\begin{document}

\title{General relativity histories theory I: \\
The spacetime character of the canonical description}
\author{ Ntina Savvidou\thanks{ntina@imperial.ac.uk}\\
{\small Theoretical Physics Group, The Blackett Laboratory,} \\
{\small Imperial College, SW7 2BZ, London, UK} \\ }

\maketitle

\begin{abstract}
The problem of time in canonical quantum gravity is related to the
fact  that the canonical description is based on the prior choice
of a spacelike foliation, hence making a   reference to a
spacetime metric. However, the metric is expected to be a
dynamical, fluctuating quantity in quantum gravity.
 We show how this problem can be
solved in the histories formulation of general relativity. We
implement the 3+1 decomposition using  metric-dependent foliations
which remain spacelike with respect to all possible Lorentzian
metrics. This allows us to find an explicit relation of
 covariant and canonical quantities which preserves the
spacetime character of the canonical description. In this new
construction, we also have a coexistence of the spacetime
diffeomorphisms group, $ \rm{Diff}(M)$, and the Dirac algebra of
constraints.
\end{abstract}

\renewcommand {\thesection}{\arabic{section}}
 \renewcommand {\theequation}{\thesection.\arabic{equation}}
\let \ssection = \section
\renewcommand{\section}{\setcounter{equation}{0} \ssection}

\pagebreak

\section{Introduction}

\subsection{On the quantisation of a Diff(M)-invariant theory of
gravitation.}

One of the major approaches to the quantisation of gravity is the
canonical one, either in its original form---involving
geometrodynamic variables \cite{ADM}---or in terms of the loop
variables (see \cite{loops} for a review), introduced via the
connection formulation of canonical general relativity
\cite{Ash86}. The canonical connection formulation of the general
quantisation  programme starts from the Hamiltonian description of
general relativity and seeks to implement some version of the
general theory of canonically quantising systems with first class
constraints.

Canonical quantisation involves the identification of a Hilbert
space on which the canonical commutation relations---or some other
appropriate algebraic structure---can be implemented, thereby
defining the kinematical variables of quantum gravity. The Hilbert
space is chosen to allow the representation of the constraints of
the Hamiltonian description in terms of self-adjoint operators,
preserving the classical Dirac algebra of constraints. Then, one
has to find the zero eigenspace of the constraint operators, in
order to define the physical Hilbert space. This is the scope of
the original Dirac quantisation of constrained systems: variations
are usually employed in the case of gravity (or special models),
because the constraint operators are not expected to have a
discrete spectrum.

In the canonical quantisation scheme, much of the research has
focused on the technical problems of constructing the Hilbert
space, writing  the constraint operators, and finding their
spectrum. However,  the canonical formalism encounters serious
problems even at the first stage of implementation. In particular,
the fact that general relativity is a generally covariant theory
raises grave doubts about the conceptual adequacy of the canonical
method of quantisation---at least in its present form.

This last remark is highlighted by the fact that the equations of
general relativity are covariant with respect to the action of the
diffeomorphisms group ${\rm Diff}(M)$, of the spacetime manifold
$M$. Although the invariance under the action of ${\rm Diff}(M)$
is particularly relevant to the notion of an `observable', it does
not pose great difficulties in the classical theory, since once
the equations of motion are solved the Lorentzian metric on $M$
can be used to implement concepts like causality and spacelike
separation.

However, in quantum theory such notions are lost, because the
geometry of spacetime is expected to be subject to quantum
fluctuations. This creates problems even at the first step of the
quantisation procedure, namely the definition of the canonical
commutation relations. The canonical commutation relations are
defined on a `spacelike' surface: however, a surface is spacelike
with respect to some particular spacetime metric $g$, which is
itself a quantum observable that is expected to fluctuate.

The prior definability of the canonical commutation relations is
not merely a mathematical requirement. In a generic quantum field
theory the canonical commutation relations implement the principle
of {\em microcausality}: namely that field observables that are
defined in spacelike separated regions commute. However, if the
notion of spacelikeness is also dynamical, it is not clear in what
way this relation will persist.

Moreover, a spacelike foliation is necessary for the
implementation of the $3+1$ decomposition and the definition of
the Hamiltonian. Again we are faced with the question of how to
reconcile  the requirement of spacelikeness  with the expectation
that different metrics will take part in the quantum description.
This problem needs to be addressed if the canonical quantum theory
is to have a spacetime character, {\i.e.} if the quantum true
degrees of freedom are to correspond to a {\em Lorentzian}
four-metric.

Even more, one may question whether the predictions of the
resulting quantum theories are independent of the choice of
foliation. The Hilbert space of the quantum theory, which it is
constructed canonically, is not straightforwardly compatible with
the ${\rm Diff}(M)$ symmetry. In the canonical theory, the
symmetry group is the one generated by the Dirac algebra of
constraints, which is mathematically distinct from the ${\rm
Diff}(M)$ group. In effect, different choices of foliation lead
{\em a priori} to different quantum theories, and there is
absolutely no guarantee that these quantum theories are unitarily
equivalent (or physically equivalent in some other generalised
sense). The canonical description cannot provide an  answer to
these questions, because once the foliation is employed for the
3+1 decomposition, its effect  is lost, and there is no way of
relating the predictions corresponding to different foliations.

These are serious problems, which challenge the validity of the
canonical approach towards the description of a generally
covariant theory of quantum gravity.

Finally, the problem which is perhaps most well known, is the {\em
problem of time} (for a review see \cite{I92, Kuc91}). The
Hamiltonian of general relativity is a combination of the first
class constraints, hence it vanishes on the reduced state space.
It is expected also to vanish on the physical Hilbert space
constructed in the quantisation scheme. This means that there is
no notion of time evolution in the space of true degrees of
freedom. More than that, the notion of time as causal ordering
seems to be lost.

In contrast, the tensorial expressions of the equations of motion
are ${\rm Diff}(M)$-invariant in the Lagrangian formalism. It is
not surprising then that Dirac\cite{Dirac} attempted to write a
Lagrangian quantum action functional analogue for general
relativity; his results led to the well known path integrals
techniques. However, path integrals cannot be well defined for
general relativity. Moreover, path-integral techniques do not
provide a full description of the quantum theory and need to be
supplemented with the introduction of Hilbert space objects---and
hence a canonical description---in order to make physical
predictions.

It seems very natural, therefore, to wish for a theory that
combines the virtues of both formalisms: the Lagrangian, and the
Hamiltonian.

\subsection{On the dual spacetime-canonical nature of histories
formalism.}

The consistent-histories approach to quantum theory was initiated
by Griffiths, Omn\'es\cite{CoHis1}, Gell-Mann and Hartle
\cite{CoHis2}. A history is defined as a sequence of time-ordered
propositions about properties of a physical system, each of which
can be represented, as usual, by a projection operator. In normal
quantum theory it is not possible to assign a probability measure
to the set of all histories; however, when a certain `decoherence
condition' is satisfied by a set of histories, the elements of
this set {\em can\/} be given probabilities. The probability
information of the theory is encoded in the decoherence
functional---a complex function of pairs of histories.

Isham and Linden proposed that propositions about the histories of
a system should be represented by projection operators on a new,
`history' Hilbert space \cite{Ish94,IL94,IL95}. An important way
of understanding the history Hilbert space ${\mathcal V}$ is in
terms of the representations of the `history group'---in
elementary systems this is the history analogue of the canonical
group. For example, for the simple case of a point particle moving
on a line, the history group for a continuous time parameter $t$
is described by the history commutation relations
 \begin{eqnarray}
 {[}x_t,x_{t'}]&=&0 = {[}p_t,p_{t'}]  \label{HA1} \\
 {[}x_t,p_{t'} ] &=& i \hbar \delta(t-t') , \label{HA3}
    \end{eqnarray}
 where the spectral projectors of the (Schr\"odinger picture)
 operators $x_t$ and $p_t$ represent the values of position and
 momentum respectively at time $t$.


 This particular history algebra is equivalent to the algebra of a
 1+1-dimensional quantum field theory, and hence techniques from
 quantum field theory (for example, for handling the problem of the
 existence of many inequivalent representations of the algebra in
 Eqs.\ (\ref{HA1}--\ref{HA3})) can be used in the study of the
 history Hilbert space. This was done successfully in
 \cite{ILSS98}, where we showed that the physically appropriate
 representation can be uniquely constructed by demanding the
 existence of a time-averaged Hamiltonian operator $H:=\int
 dt\,H_t$.

A significant result emerged from the study of continuous-time
transformations. Namely, that there exist {\em two\/} distinct
generators of time transformation \cite{Sav99}. One refers to time
as an ordering parameter ($t$-label in Eqs.\
(\ref{HA1}--\ref{HA3})), which is related to the causal structure
and the kinematics of a theory. The other generator refers to time
as it appears in the implementation of dynamical laws (the label
$s$ in the `history Heisenberg picture' operator $\;\;
x_t(s):=e^{isH}x_t e^{-isH}$), and it is related to the
Hamiltonian evolution and the dynamics of a theory.

Most importantly: for any specific physical system these two
transformations are intertwined by the definition of the action
operator---a quantum analogue of the classical action functional.
Hence, the definition of these two distinct operators implementing
time transformations signifies the {\em distinction between the
kinematics and the dynamics of the theory\/}. This distinction and
the corresponding definitions are also valid for classical
histories.

One of the most important consequences of the histories approach
is  that a combined spacetime-canonical formalism emerges. The
richer temporal structure of a history theory allows the
simultaneous description of both spacetime and canonical objects.

In a preliminary study \cite{SavGR01}, we presented a history
version of general relativity,  which demonstrates a new relation
between the group structures, associated to the Lagrangian and
Hamiltonian approaches. In particular, we showed that in this
histories version of canonical general relativity there exists a
representation of the spacetime diffeomorphisms group ${\rm
Diff}(M)$, together with a history analogue of the Dirac algebra
of constraints.

However, various important issues arose. First, the history
canonical algebra depends on the choice of a Lorentzian foliation.
Hence, a natural question is the degree to which physical results
depend upon this choice. For each choice, the solutions to the
equations of motion  enable us to construct different 4-metrics.
If different descriptions are to be equivalent, two distinct
4-metrics should be related by a spacetime diffeomorphism. We
should therefore establish that the action of the spacetime
diffeomorphisms group intertwines between the constructions
associated with the different choices of the foliation. This
involves the comparison of history state space associated with
arbitrary choices of foliation.

Second, and perhaps more important, we need to question the notion
of a spacelike foliation itself. Since the spacetime causal
structure is a dynamical object, the notion of a foliation being
{\em spacelike\/} has meaning only {\em after\/} the solution to
the classical equations of motion has been selected. However, in
the histories description we do not use just  a single solution of
the classical equations of motion (indeed, many of the possible
histories are not solutions at all), and in these circumstances
the notion of a `spacelike' foliation loses its meaning.

These are some of the deepest issues not only of history theories
but of any canonical approach to gravity. Hence, they inevitably
require a significant reworking of the theory. In the present
paper, we successfully address these issues by focusing on a
crucial point of the histories formalism: the connection between
the covariant and the canonical description of histories general
relativity.

The key idea of this new construction is the introduction of the
notion of a {\em metric-dependent foliation\/}. Specifically, we
choose foliations  that are functionals of the four-metric $g$,
and that are required to be spacelike with respect to $g$. In
particular, under the action of a spacetime diffeomorphism, a
foliation that is spacelike will preserve this character as both
the embedding and the metric will transform together. This is a
simple idea, but it works very successfully for the histories
formalism, and allows us to include all different choices of
foliation  in studying the foliation dependence of the history
canonical algebra.

The plan of the paper is as follows. In section 2 we write a brief
summary of the histories formalism. In section 3.1 we present the
spacetime description of histories general relativity, and we
write the representation of the spacetime diffeomorphisms ${\rm
Diff}(M)$. In section 3.2 we give a detailed presentation of the
relation between the spacetime and the canonical descriptions. We
emphasise the difference between foliations that do not have a
metric-dependence, and those that do, and we use the latter to
construct the explicit relation between the covariant history
space $\Pi^{cov}$ and the canonical history space $\Pi^{can}$.
Next, we write the symplectic forms for both descriptions, and we
show that they are equivalent.

In section 3.3, we present the canonical description of histories
general relativity. We write the  Dirac algebra of constraints,
and we show that in the histories theory of metric-dependent
foliations, there exist representations of both the group of
spacetime diffeomorphisms and of the Dirac algebra of constraints.

\section{Background}

\paragraph{Temporal Structure of HPO histories theory.} Although the
histories programme originated from the consistent histories
theory, it was developed with an emphasis on the  `temporal' logic
of the theory \cite{Ish94}. However, the HPO (`Histories
Projection Operator') theory takes a completely different turn
once the concept of time is introduced in a new way \cite{Sav99}.

In the consistent histories formalism,  a history $ \alpha = (
\hat{\alpha}_{t_{1}},\hat{\alpha}_{t_{2}},\ldots,
\hat{\alpha}_{t_{n}})$ is defined to be a collection of projection
operators $\hat\alpha_{t_i}$, $i=1,2,\ldots,n$, each of which
represents a property of the system at the single time $t_i$.
Therefore, the emphasis is placed on histories, rather than
properties at a single time, which in turn gives rise to the
possibility of generalized histories with novel concepts of time.

The HPO approach, places particular emphasis on temporal logic.
This is achieved by representing the history $\alpha$ as the
operator $\hat\alpha :=\hat{\alpha}_{t_{1}}\otimes\hat{\alpha}
_{t_{2}}\otimes\cdots \otimes \hat{\alpha}_{t_{n}}$ which is a
genuine {\em projection\/} operator on the tensor product $
\otimes_{i=1}^n {\mathcal{H}}_{t_i} $ of copies of the standard
Hilbert space $\mathcal{H}$. Note that to use this construction in
any type of field theory requires an extension to a continuous
time label, and hence to an appropriate definition of the
continuous tensor product $\otimes_{t\in\mathR}{\mathcal{H}}_t$
\cite{IL95, An01b}.

 A central feature of the  histories theory is
the development of the novel temporal structure \cite{Sav99},
namely the existence of {\em two\/} distinct types of time
transformation.

\paragraph{Classical Histories} The space of classical histories $ \Pi = \{
\gamma \mid \gamma : \mathR \rightarrow \Gamma \}$ is the set of
all smooth paths on the classical state space $\Gamma$. It can be
equipped with a natural symplectic structure, which gives rise to
 Poisson brackets. For the simple case of a particle on a line, we
 have
\begin{eqnarray}
    \{x_t\, , x_{t^{\prime}} \}_{\Pi} &=& 0\\
    \{p_t\, , p_{t^{\prime}} \}_{\Pi} &=& 0\\
  \{x_t\, , p_{t^{\prime}} \}_{\Pi} &=& \delta (t-t^{\prime})
  \end{eqnarray}
where
\begin{eqnarray}
x_t : \Pi & \rightarrow & \mathR  \\
\gamma & \mapsto & x_t (\gamma) := x( \gamma (t))
\end{eqnarray}
and similarly for $ p_t $.

The classical analogue of the Liouville operator is defined as
 \begin{equation}
  V(\gamma) := \int_{-\infty}^\infty \! dt\, p_t\, \dot{x}_t ,
 \end{equation}
and the Hamiltonian ({\em i.e.}, time-averaged energy) function
$H$ is defined as
\begin{equation}
  H ( \gamma ):= \int_{-\infty}^{\infty}\!dt \,H_t(x_t,p_t)
\end{equation}
where $H_t$ is the Hamiltonian that is associated with the copy
${\Gamma}_t$ of the normal classical state space with the same
time label $t$.

The temporal structure leads to the histories analogue of the
classical equations of motion
\begin{equation}
\{ F , V \}_{\Pi}\, (\gamma_{cl}) = \{ F , H \}_{\Pi}\,
(\gamma_{cl})
\end{equation}
where $F$ is any function on $\Pi$, and where the path
$\gamma_{cl}$ is a solution of the equations of motion.

A crucial result is that the history equivalent of the classical
equations of motion is given by the following condition that holds
for \textit{all} functions $F$ on $\Pi$ when $\gamma_{cl}$ is a
classical solution:
\begin{equation}
\{F , S\}_{\Pi}\, (\gamma_{cl}) = 0,
\end{equation}
where
\begin{equation}
S( \gamma ) := \int_{-\infty}^{\infty}\!dt \, (p_t\dot{x_t} -
H_t(x_t,p_t))=\, V( \gamma )- H( \gamma ) \label{Sclass}
\end{equation}
is the classical analogue of the action operator. This is the
history analogue of the least action principle \cite{Sav99}.

The temporal structure of HPO histories enables us to treat
parameterised systems in such a way that the problem of time does
not arise \cite{SA00}. Indeed, histories keep their intrinsic
temporality after the implementation of the constraint: thus there
is no uncertainty about the temporal-ordering properties of the
physical system.

\paragraph{Quantum histories.}
In the corresponding quantum theory, the Hamiltonian operator $H$
and the `Liouville' operator $V$ are the generators of the two
types of time transformation \cite{Sav99}. Specifically, the
Hamiltonian $H$ is the generator of the unitary time evolution
with respect to the `internal' time label $s$; this has no effect
on the  time label $t$. On the other hand, the Liouville operator
$V$---defined in analogy to the kinematical part of the classical
action functional---generates time translations along the $t$-time
axis without affecting the $s$-label.

We can define the \textit{action} operator $S$ as a quantum
analogue of the classical action functional Eq. (\ref{Sclass}). It
transpires that the action operator $S$ generates \textit{both}
types of time transformation, and in this sense it is the
generator of {\em physical} time translations in the histories
formalism.

The time transformations generated by the action operator $S$
resemble the canonical transformations generated by the
Hamilton-Jacobi action functional. Indeed, there is an interesting
relation between the definition of $S$ and the well-known work by
Dirac on the Lagrangian theory for quantum mechanics
\cite{Dirac,Sav99}. In particular, motivated by the fact
that---contrary to the Hamiltonian method---the Lagrangian method
can be expressed relativistically (on account of the action
function being a relativistic invariant), Dirac tried to take over
the general $\it{ideas}$ of the classical Lagrangian theory,
albeit not the equations of the Lagrangian theory {\em per se\/}.

\section{Histories General Relativity}

\subsection{Spacetime description of histories general relativity theory}

In order to apply the histories theory to general relativity we
start with the description of spacetime quantities. We consider a
four-dimensional manifold $M$, which has the topology $\mathR
\times \Sigma$. The history space is defined as $\Pi^{cov} =
T^{*}{\rm LRiem}(M)$, where ${\rm LRiem}(M)$ is the space of all
Lorentzian four-metrics ${g_{\mu\nu}}$, and $T^{*}{\rm LRiem}(M)$
is its cotangent bundle. Hence, the history space $\Pi^{cov}$ for
general relativity is the space of all histories $(g_{\mu\nu},
\pi^{\mu\nu})$.

The history space $\Pi^{cov}$ is equipped with the symplectic form
\begin{equation}
 \Omega = \int \!d^4X \, \delta \pi^{\mu\nu}(X) \wedge \delta
 g_{\mu\nu}(X) \,, \label{omega}
\end{equation}
where $X$ is a point in the spacetime $M$, and where
$g_{\mu\nu}(X)$ is a four-metric that belongs to the space of
Lorentzian metrics $L {\rm Riem}(M)$, and $\pi^{\mu\nu}(X)$ is the
conjugate variable.

The symplectic structure Eq.\ (\ref{omega}) generates the
following covariant Poisson brackets algebra, on the history space
$\Pi^{cov}$
\begin{eqnarray}
\{g_{\mu\nu}(X)\,, \,g_{\alpha\beta}(X^{\prime})\}&=& 0 \label{covgg}\\
\{\pi^{\mu\nu}(X)\,, \,\pi^{\alpha\beta}(X^{\prime})\} &=& 0
\label{covpipi} \\
\{g_{\mu\nu}(X)\,, \,\pi^{\alpha\beta}(X^{\prime})\} &=&
\delta_{(\mu\nu )}^{\alpha\beta} \,\delta^4 (X, X^{\prime}) ,
\label{covgpi}
\end{eqnarray}
 where  ${{\delta}_{(\mu\nu)}}^{\alpha\beta} :=
\frac{1}{2}({\delta_\mu}^\alpha {\delta_\nu}^\beta +
{\delta_\mu}^\beta {\delta_\nu}^\alpha )$.

\subsubsection{The representation of the group ${\rm Diff}(M)$}
The critical observation now is that we can write a representation
of the spacetime diffeomorphisms group ${\rm Diff}(M)$ on the
history space $\Pi^{cov}$.

In previous applications of the histories formalism we defined the
Liouville function $V$ as the generator of time translations with
respect to the external time $t$ that appears as a kinematical
ordering parameter \cite{Sav99, SA00, Sav01}. In the present case
we  define, by analogy, the Liouville function $V_W$ associated
with any vector field $W$ on $M$ as
\begin{equation}
V_W:=\int \!d^4X \,\pi^{\mu\nu}(X)\,{\cal L}_W g_{\mu\nu}(X)
\label{Vw}
\end{equation}
where ${\cal L}_W$ denotes the Lie derivative with respect to $W$.
This is the analogue of the expression that is used in the normal
canonical theory for the representations of spatial
diffeomorphisms.

The fundamental result is that these generalised Liouville
functions $V_W$, defined for any vector field $W$ as in Eq.\
(\ref{Vw}),  satisfy the Lie algebra of the {\em spacetime
diffeomorphisms group\/} ${\rm Diff}(M)$:
\begin{equation}
\{\, V_{W_1}\,, V_{W_2}\,\} = V_{ [ W_1 , W_2 ]},
\end{equation}
where $[ W_1 , W_2 ]$ is the Lie bracket between vector fields
$W_1$ and $W_2$ on the manifold $M$.

We note here that the functional $V_{W}$ acts upon the basic
variables of the theory as an infinitesimal diffeomorphism:
\begin{eqnarray}
 \{\, g_{\mu\nu}(X)\,, V_W \,\} &=& {\cal L}_W g_{\mu\nu}(X)   \\
 \{\, \pi^{\mu\nu}(X)\,, V_W \,\} &=& {\cal L}_W \pi^{\mu\nu}(X) .
\end{eqnarray}

\subsection{Relation between spacetime and
canonical descriptions}

Next,  we show how the histories Dirac algebra of constraints
appears in the history space $\Pi^{cov}$. For this reason we study
the relation of  the covariant (spacetime) description with its
canonical (evolutionary) counterpart.

We  introduce a $3+1$ foliation of the spacetime $M$, which is
spacelike with respect to a Lorentzian metric $g$, in order to
construct a $3 + 1$ description of the theory. However, a key
feature of the present construction is that this foliation is
required to be {\em four-metric dependent\/} in order to address
the key issue of requiring all the different choices of foliation
to be spacelike. We will then show the relation between the
covariant history space $\Pi^{cov}$ and its canonical counterpart
$ \Pi^{can}$.

In the appendix A we have collected some mathematical definitions
that are necessary for the understanding of the connection between
the covariant, and the $3+1$, formulations of the theory.

\subsubsection{Foliations not depending on four-metric $g$}

We consider the spacetime manifold $M = \mathR\times\Sigma$ and
the space ${\rm Fol}(M)$ of all foliations of $M$ that are
spacelike with respect to at least one Lorentzian metric. For each
specific Lorentzian metric $g$ we choose a spacelike foliation,
${\cal E}:\mathR\times\Sigma\rightarrow M$ with an associated
family of spacelike embeddings ${\cal E}_t:\Sigma\rightarrow M$,
$t\in\mathR$. We then define the pull-back of $g_{\mu\nu}$ to
$\mathR\times\Sigma$ as ${\cal E}^*g$. We also wish to pull-back
the conjugate variable $\pi^{\alpha \beta}$ to
$\mathR\times\Sigma$. For this purpose, we lower the indices and
define the field $\pi_{\alpha \beta}(X) =
g_{\gamma\alpha}(X)\,g_{\zeta\beta}(X)\, \pi^{\gamma\zeta}(X)$,
which is  a $(0,2)$ tensor that can be pull-backed on $\Sigma$ in
the usual way.

Using the Poisson bracket equations \ (\ref{covgg}--\ref{covgpi}),
we get the relations
\begin{eqnarray}
\{g_{\mu\nu}(X)\,, \,g_{\alpha\beta}(X^{\prime})\}&=& 0
\label{covigg}  \\
\{\pi_{\mu\nu}(X)\,, \,\pi_{\alpha\beta}(X^{\prime})\} &=& 0
\label{covipipi} \\
\{g_{\mu\nu}(X)\,, \,\pi_{\alpha\beta}(X^{\prime})\} &=&
{g_{(\mu\alpha}}\,{g_{\nu)\beta}(X)} \;\delta^4 \!(X,\,
X^{\prime}) \label{covigpi}
\end{eqnarray}
where $g_{(\mu\alpha}g_{\nu)\beta}(X):= \frac{1}{2}
(g_{\mu\alpha}(X)\,g_{\nu\beta}(X) +
g_{\nu\alpha}(X)\,g_{\mu\beta}(X))$.

We define the deformation vector Eq.\ (\ref{f4}), that is uniquely
selected by the choice of this one-parameter family of embeddings
of $\Sigma$ in $M$. This family of embeddings also allows the
selection of a coordinate system common to all the embedded
three-surfaces in the sense that the coordinate defined on the
reference three-surface $\Sigma$ is shared by all of them.

Next we define the spatial parts of the pull-back of
$g_{\mu\nu}(X)$ to $\mathR\times\Sigma$ by ${\cal E}$ as
\begin{equation}
h_{ij}(t,\underline{x}) := {\cal{E}}^{\mu}_{,i}(t,\underline{x})\,
{\cal{E}}^{\nu}_{,j}(t,\underline{x})\,
g_{\mu\nu}({\cal{E}}(t,\underline{x})) \label{Pullbackh}
\end{equation}
where ${\cal{E}}^{\mu}_{,i}(t,\underline{x}):={\partial}_i
({\cal{E}}^{\mu}(t,\underline{x}))$.

The choice of a foliation $\cal{E}$, spacelike with respect to a
specific metric $g$, uniquely defines the lapse function $N$ and
shift vector $N_i$ of the $3 + 1$ decomposition of the
four-metric, as
\begin{eqnarray}
 N(t,\underline{x}) &=& {\dot{\cal{E}}}^{\mu}\, n_{\mu}({\cal{E}}
 (t,\underline{x})) \label{lapse} \\
 N_{i}(t,\underline{x}) &=& {\cal{E}}^{\mu}_{,i}\, {\cal{E}}^{\nu}_{,j}\,
 {\dot{\cal{E}}}^{\nu} \, g_{\mu\nu}({\cal{E}}(t,\underline{x})),
 \label{shift}
\end{eqnarray}
where $n_{\mu}$ is the unit, timelike vector, normal to the
foliation, and
\begin{equation}
 g^{\mu\nu} =
 -\frac{1}{{N}^2}\,{\dot{\cal{E}}}^{\mu}\,{\dot{\cal{E}}}^{\nu} +
 ({\dot{\cal{E}}}^{\mu}\,{\cal{E}}^{\nu}_{,i} +
{\cal{E}}^{\mu}_{,i} \,{\dot{\cal{E}}}^{\nu})\, \frac{N_{i}}{N^2}
+ (h^{ij} - \frac{N_{i} N_{j}}{N^2})\,
{\cal{E}}^{\mu}_{,i}\,{\cal{E}}^{\nu}_{,j}. \label{g:3+1}
\end{equation}

\bigskip

\subsubsection{Foliations depending on four-metric $g$}

We have showed so far that, for a fixed metric $g$ we can choose
the foliation to be spacelike in the sense that $t\mapsto
h_{ij}(t,\underline{x})$ is a path in the space of Riemannian
metrics on $\Sigma$. For an appropriate topology on $L{\rm
Riem}(M)$, this spacelike character will be maintained for some
open neighborhood of the Lorentzian metric $g$. However, this
foliation  fails to be spacelike for most other Lorentzian metrics
on $M$. This feature is not important at the level of the
classical theory, because we only consider the four-metric, which
is the solution of the equations of motion; however it can be
expected to be a non-trivial issue in the quantum theory.

When we consider a fixed foliation for all four-metrics, there
will be  metrics $g \in LRiem(M)$, for which some of the pullbacks
${\cal E}_t^* g$, $t\in\mathR$, will not be a Riemannian
three-metric on $\Sigma$. In other words, the pull-back space
${\cal E}_t^* LRiem(M)$ does not coincide with the space of
Riemannian metrics $ Riem( \Sigma )$, which is the space of the
canonical description of general relativity.

We want to place special emphasis on this point: it reflects one
of the major conceptual problems of the canonical description of
gravity, and it is directly related to the problem of time in
quantum gravity. As explained in the Introduction, a general
relativity canonical description involves the choice of a specific
spacelike foliation; however, in a theory where the metric is a
non-deterministic dynamical variable---as it is expected in
quantum gravity---the notion of `spacelike' has no a priori
meaning.

In order to address this important issue we introduce the notion
of a metric-dependent foliation.

To this end, for each $g \in LRiem(M)$ we choose a spacelike
foliation ${\cal E}[g]$.
For a given Lorentzian metric $g$, we use the foliation
${\cal{E}}[g]$ to split $g$ with respect to the Riemannian
three-metric $h_{ij}$, the lapse function $N$ and the shift vector
$N^i$ as
\begin{eqnarray}
 h_{ij}(t,\underline{x};g] &:=&
 {\cal{E}}^{\mu}_{,i}(t,\underline{x};g]\,
 {\cal{E}}^{\nu}_{,j}(t,\underline{x};g]\,
 g_{\mu\nu}({\cal{E}}(t,\underline{x};g]) \label{Pullbackh[g]} \\
 N_{i}(t,\underline{x};g] &:=&
 {\cal{E}}^{\mu}_{,i}(t,\underline{x};g]\,
 {\dot{\cal{E}}}^{\nu}(t,\underline{x};g]\, g_{\mu\nu}({\cal{E}}
 (t,\underline{x};g])\label{shift[g]} \\
 -N^{2}(t,\underline{x};g] &:=&
 {\dot{\cal{E}}}^{\mu}(t,\underline{x};g]\,
 {\dot{\cal{E}}}^{\nu}(t,\underline{x};g]\,
 g_{\mu\nu}({\cal{E}}(t,\underline{x};g]) - N_{i} N^{i} (t,
 \underline{x})
 \label{lapse[g]}
\end{eqnarray}

\subsubsection{The relation between the covariant history space
$\Pi^{cov}$ and the history space of the canonical description
$\Pi^{can}$}

We have showed that the history space of the spacetime description
of histories general relativity $\Pi^{cov} = T^{*} LRiem(M)$ is
equipped with the symplectic structure characterised by the
symplectic form Eq.\ (\ref{omega}). In order to relate $\Pi^{cov}$
with its canonical counterpart, it suffices to write its
symplectic form $\Omega$ in terms of the canonical variables
$h_{ij}$, $N_i$ and $N$---that enter in the $3+1$ decomposition of
the Lorentzian metric $g$---and their corresponding conjugate
momenta.

To this end, starting from Eq.\ (\ref{g:3+1}), we have
\begin{eqnarray}
 \delta\! g^{\mu\nu}\!\! =\!\! [\delta(\frac{-1}{N^2})
 {\dot{\cal{E}}}^{\mu}\!
 {\dot{\cal{E}}}^{\nu} + \delta(\frac{N_{i}}{N^2})
 ({\dot{\cal{E}}}^{\mu}\!{\cal{E}}^{\nu}_{,i} + {\cal{E}}^{\mu}_{,i}
{\dot{\cal{E}}}^{\nu}) + \delta
h^{ij}{\cal{E}}^{\mu}_{,i}\!{\cal{E}}^{\nu}_{,j} -
\delta(\frac{N_{i} N^{j}}{N^2})
{\cal{E}}^{\mu}_{,i}\!{\cal{E}}^{\nu}_{,j} ] \nonumber \\
 \vspace*{3cm}+ [\frac{-1}{N^2}\delta({\dot{\cal{E}}}^{\mu}\!
 {\dot{\cal{E}}}^{\nu}) + \frac{N_{i}}{N^2}
 (\delta({\cal{E}}^{\mu}_{,i}\!{\dot{\cal{E}}}^{\nu}) +
 \delta({\dot{\cal{E}}}^{\mu}\!{\cal{E}}^{\nu}_{,i}))  +
 (h^{ij}
 - \frac{N^{i} N^{j}}{N^2})\,
 \delta({\cal{E}}^{\mu}_{,i}\!{\cal{E}}^{\nu}_{,j})]. \label{deltag}
\end{eqnarray}
We note that the first bracket of the right-hand side of Eq.\
(\ref{deltag}) corresponds to the variation of $g$ (in terms of
$\delta N_i$, $\delta N$, and $\delta h^{ij}$) that would occur if
the foliation was not metric-dependent; we will denote this term
as $A^{\mu\nu}$. The terms within the second bracket arise from
the fact that the foliation {\em is\/} metric dependent. These
extra terms can be written in the form
$(B^{\mu\nu}_{\rho\sigma}+C^{\mu\nu}_{\rho\sigma}+
D^{\mu\nu}_{\rho\sigma})\delta g^{\rho\sigma}$ where
\begin{eqnarray}
 B^{\mu\nu}_{\rho\sigma}(X,X^{\prime})\!\!\!&:=&\!\! (\frac{-2}
 {N^2}
 {\dot{\cal{E}}}^{\nu}(X;g)
 + \frac{2 N^i}{N^2} {\cal{E}}^{\nu}_{,i}(X;g))
 \frac{\delta {\dot{\cal{E}}}^{\mu}(X;g)}{\delta
 g^{\rho\sigma}(X^{\prime})}     \label{B} \\
 C^{\mu\nu}_{\rho\sigma}(X,X^{\prime})\!\!\!&:=&\!\!\!
 (\frac{2 N^i}{N^2}
 {\dot{\cal{E}}}^{\nu}(X;g)
 + (h^{ij}\! -\! \frac{N^i N^j}{N^2}
 {\cal{E}}^{\nu}_{,j}(X;g))
 \frac{\delta {{\cal{E}}^{\mu}_{,i}}(X;g)}{\delta
 g^{\rho\sigma}({X^{\prime}})} \label{C}   \\
 D^{\mu\nu}_{\rho\sigma}(X,X^{\prime})\!\!\!&:=&\!\! (\frac{N^i
 N^j}{N^2})
 {\cal{E}}^{\mu}_{,i}(X;g))
 \frac{\delta {\cal{E}}^{\nu}_{,j}(X;g)}{\delta
 g^{\rho\sigma}(X^{\prime})} \label{D}
\end{eqnarray}
so that we have
\begin{equation}
\delta g^{\mu\nu}=A^{\mu\nu}+
(B^{\mu\nu}_{\rho\sigma}+C^{\mu\nu}_{\rho\sigma}+
D^{\mu\nu}_{\rho\sigma})\delta g^{\rho\sigma}
\end{equation}
which can be rewritten as
\begin{equation}
    A^{\mu\nu}=E^{\mu\nu}_{\rho\sigma}\delta g^{\mu\nu}
    \label{A=Eg}
\end{equation}
where
\begin{equation}
 E^{\mu\nu}_{\rho\sigma}(X,X^{\prime})\!\!\! := \!\!(1-B-C-D)^{\mu\nu}_
 {\rho\sigma} (X,X^{\prime}). \label{E}
\end{equation}

For Eq.\ \ref{A=Eg} to be meaningful, it is necessary that that
the inverse of the tensor $E^{\mu\nu}_{\rho\sigma}$ exists. This
holds for a metric-independent $\cal E$  since then $B=C=D=0$, and
the condition will continue to be satisfied for small values of
the functional derivative $\delta{\cal E}^\mu/\delta
g^{\rho\sigma}$. There is a prima facie possibility that $E$ could
become non-invertible for sufficiently large values of
$\delta{\cal E}^\mu/\delta g^{\rho\sigma}$, but this is part of
the general question of the overall global structure of the
history symplectic space and is not something with which we shall
concern ourselves in the present paper.

With this proviso, detailed calculations show that the symplectic
form $\Omega$ can be written in the equivalent canonical form,
with respect to a chosen element, $\cal E$, of ${\rm Fol}_g(M)$,
as
\begin{eqnarray}
 \Omega &=& \int \!d^4X \, \delta \pi^{\mu\nu} \wedge \delta
 g_{\mu\nu} =  -\int \!d^4X \, \delta \pi_{\mu\nu} \wedge \delta
 g^{\mu\nu}  \label{omega3+1} \\ \nonumber
 &=& \int\!d^3x\,dt (\delta \tilde{\pi}^{ij}\wedge\delta h_{ij} +
 \delta {\tilde{p}} \wedge\delta N
 + \delta \tilde{p_{i}}\wedge \delta N^{i}), \label{omegacan}
\end{eqnarray}
where
\begin{eqnarray}
\tilde{\pi}^{ij} := \! \! K(t,\underline{x})
(E^{-1\top}\pi)_{\mu\nu}
 h^{ik} \, h^{jl} \,{\cal{E}}^{\mu}_{,k} \,{\cal{E}}^{\nu}_{,l}
  \label{gpijtilde}  \\
 \tilde{p} :=\! \!- K(t, \underline{x}) \!{(E^{-1\top}\!\pi)}
\!_{\mu\nu} \frac{2}{N^3}
({\dot{\cal{E}}}^{\mu}{\dot{\cal{E}}}^{\nu}\!\!
 -\!\!2 N^{i}({\cal{E}}^{\mu}_{,i} {\dot{\cal{E}}}^{\nu}
 \!+\!
 {\dot{\cal{E}}}^{\mu} {\cal{E}}^{\nu}_{,i})\! + \! N^{i} N^{j}
 {\cal{E}}^{\mu}_{,i}{\cal{E}}^{\nu}_{,j}) &&   \label{gptilde}  \\
 \tilde{p}_i := - K(t, \underline{x}) \!{(E^{-1\top}\!\pi)}\!_{\mu\nu}
 ({\cal{E}}^{\mu}_{,i} {\dot{\cal{E}}}^{\nu} +
 {\dot{\cal{E}}}^{\mu} {\cal{E}}^{\nu}_{,i} - N^{j}
 ({\cal{E}}^{\mu}_{,i}{\cal{E}}^{\nu}_{,j} +
 {\cal{E}}^{\mu}_{,j}{\cal{E}}^{\nu}_{,i})).&& \label{gpitilde}
\end{eqnarray}
Here $K(t,x)$ is the determinant of the transformation from the
$X$ to the $(t,x)$ variables. Given that the volume form reads
\begin{eqnarray}
\sqrt{-g} \, dX^0 \wedge dX^1 \wedge dX^2 \wedge dX^3 = N
\sqrt{\tilde{h}} \, dt \wedge dx^1 \wedge dx^2 \wedge dx^3,
\end{eqnarray}
we identify $K(t,x)$ as
\begin{eqnarray}
K(t, \underline{x}) = \frac{N(t, \underline{x})
\sqrt{\tilde{h}}(t, \underline{x})}{\sqrt{-g}({\cal E}(t,
\underline{x}))}.
\end{eqnarray}
Here $\tilde{h}$ is the determinant of the matrix $h_{ij}$. Note
that it is a density of weight 1 with respect to time as well as
spatial variables and this renders $\tilde{\pi}^{ij}, \tilde{p}_i,
\tilde{p}$ densities of weight 1 with respect to time \footnote{It
is customary in the canonical description to consider the lapse
function as a density. However, by its definition (3.17) the lapse
function is a scalar function on $M$. The reason it is considered
as density with respect to time is equation (3.35). The
determinant  $\sqrt{\tilde{h}} $ is, strictly speaking, a density
of weight 1 with respect to time, even though it is defined by
means of the spatial metric $h_{ij}$. However, in the canonical
treatment the time-density nature of $\tilde{h}$ is ignored and
for this reason the lapse function is implicity considered as
containing the weight of the temporal density.}.

 In terms of the normal vector $n^{\mu} = \frac{1}{N}
(\dot{{\cal E}}^{\mu} - N^i {\cal E}^{\mu}_i)$, the momenta
$\tilde{p}$ and $\tilde{p}_i$ are
\begin{eqnarray}
\tilde{p}\!\!&:=& \vspace*{2cm}- K(t,
\underline{x})\frac{2}{N}{(E^{-1\top}\pi)}^{\mu\nu} n_{\mu}
n_{\nu} \\
\tilde{p}_i \!\!&:=& \vspace*{2cm} - K(t, \underline{x})
\,{(E^{-1\top}\pi)}_{\mu\nu} (n^{\mu} \dot{{\cal E}}^{\nu}_i +
n^{\nu} \dot{{\cal E}}^{\mu}_i)
\end{eqnarray}

In the special case of a metric-independent foliation, we recover
the familiar definitions
\begin{eqnarray}
 \tilde{\pi}^{ij} &:=&  K(t, \underline{x})\,{\pi}_{\mu\nu}
 h^{ik}h^{jl}{\cal{E}}^{\mu}_{,k}{\cal{E}}^{\nu}_{,l}  \label{pijtilde} \\
 \tilde{p} &:=& \!\!\!- K(t, \underline{x}) \,{\pi}_{\mu\nu}
 \frac{2}{N} n^{\mu} n^{\nu}    \label{ptilde}  \\
 \tilde{p}_i &:=& \!\!- K(t, \underline{x})( \,{\pi}_{\mu\nu}
 n^{\mu} \dot{{\cal E}}^{\nu}_i +
n^{\nu} \dot{{\cal E}}^{\mu}_i). \label{pitilde}
\end{eqnarray}

To formulate the final step of the connection between the
covariant and the canonical histories spaces, we recall that, in
the histories formalism, the basic element is a history, which is
a path $t \mapsto \Gamma$. The objects
$\tilde{p}(t,\underline{x})$ and $\tilde{p}_i(t,\underline{x})$
are densities with respect to reparameterisations of the $t$
label, hence the association $t\mapsto \tilde{p}(t,\underline{x})$
does {\em not\/} correspond to a path in the space of scalar
fields on $\Sigma$. For this reason, we can use as history
canonical variables the objects $\pi^{ij}(t,\underline{x})$,
$p(t,\underline{x})$ and $p_i(t,\underline{x})$, that are scalar
functions with respect to the time variable $t$. Hence, we define
the scalar histories quantities
\begin{eqnarray}
 \pi^{ij}(t,\underline{x})&:=& \alpha(t) {\tilde{\pi}}^{ij}
(t,\underline{x})
 \label{pij} \\
  p_{i}(t,\underline{x}) &:=& \alpha(t) \tilde{p}_{i}(t,\underline{x})
   \label{Ni} \\
 p(t,\underline{x}) &:=& \alpha(t) \tilde{p}(t,\underline{x}),  \label{N}
\end{eqnarray}
where $\tilde{N}$ is defined from Eq.\ (\ref{lapse[g]}), and where
$\alpha(t)$ is some strictly positive scalar density of weight
$-1$ in the variable $t$\footnote{The three-metric $h_{ij}$, and
the conjugate lapse function $p$ and the conjugate shift vector
$N_i$ are scalar functions with respect to time.}.

Finally, the symplectic form $\Omega$ can be written in its
equivalent {\em histories} canonical form as
\begin{eqnarray}
 \Omega &=& \int \!d^4X \, \delta \pi^{\mu\nu} \wedge \delta
 g_{\mu\nu} = -\int \!d^4X \, \delta \pi_{\mu \nu} \wedge \delta
 g^{\mu\nu}  \label{omegacc} \\ \nonumber
 &=& \int\!d^3x \frac{dt}{\alpha(t)} (\delta \pi^{ij}\wedge\delta h_{ij}
+ \delta p \wedge\delta N
 + \delta p_{i}\wedge \delta N^{i}), \label{homegacan}
\end{eqnarray}

Hence, the covariant histories space $\Pi^{cov}$ is equivalent to
the canonical histories space $P^{can} = {\times}_{t} (T^{*}{\rm
Riem}({\Sigma}_{t}) \times T^{*}Vec({\Sigma}_{t}) \times
T^{*}C^{\infty}({\Sigma}_{t}))$, where ${\rm Riem}({\Sigma}_{t})$
is the space of all Riemannian three-metrics on the surface
${\Sigma}_{t}$, $Vec({\Sigma}_{t})$ is the space of all vector
fields on ${\Sigma}_{t}$, and $C^{\infty}({\Sigma}_{t})$ is the
space of all smooth scalar functions on ${\Sigma}_{t}$.

It is important to stress, once more, that this equivalence is
only possible because of the introduction of the metric-dependent
foliation. In its absence, the canonical histories do not
correspond, in general, to genuine spacetime quantities, namely
{\em Lorentzian} metrics.

\subsection{Canonical description of histories general relativity theory}
In the previous section, we presented in detail the connection
between the covariant and the canonical description of histories
general relativity. In particular, we explained the relation
between the respective histories spaces $\Pi^{cov}$ and
$\Pi^{can}$, and we properly defined the histories variables of
the canonical description, in relation to the $3+1$ decomposition
of the Lorentzian metric $g$ with respect to a metric-dependent
foliation ${\cal{E}}(X;g]$.

In this section, we will present in detail the canonical treatment
of the theory, and we will write explicitly the representation of
the Dirac algebra of constraints.

\subsubsection{Canonical treatment: basic structure}

The history space $\Pi^{can}$ of the canonical description  is a
suitable subset of the Cartesian product ${\times}_t {\Gamma}_t $
of copies of the classical general relativity state space $\Gamma
= \Gamma(\Sigma)$, labelled by a parameter $t$, with $t\in
\mathR$. Here $\Sigma$ is a fixed three-manifold.

In particular, we have showed above that $\Gamma (\Sigma) =
T^{*}{\rm Riem}(\Sigma) \times T^{*}Vec(\Sigma) \times
T^{*}C^{\infty}(\Sigma)$, where ${\rm Riem}(\Sigma)$ is the space
of Riemannian metrics on $\Sigma$; {\em i.e.}, an element of
$\Gamma (\Sigma)$ is a pair $ ( h_{ij}, \pi^{kl}, N^{i}, p_{i}, N,
p )$. A history is defined to be any smooth map $t\mapsto
(h_{ij}(t,\underline{x}),\pi^{kl}(t,\underline{x}),
N^{i}(t,\underline{x}), p_{i}(t,\underline{x}),
N(t,\underline{x}), p(t,\underline{x}))$.

The history version of the canonical Poisson brackets can be
derived from the covariant Poisson brackets Eqs.\
(\ref{covgg})--(\ref{covgpi}) as
\begin{eqnarray}
\{h_{ij} (t,\underline{x})\,, h_{kl} ( t^{\prime} ,
\underline{x}^{\prime} ) \} &=& 0 \label{GR1b}\\
\{ \pi^{ij} (t,\underline{x})\,, \pi^{kl} ( t^{\prime} ,
\underline{x}^{\prime} ) \} &=& 0 \label{GR2b} \\
\{ h_{ij} (t,\underline{x})\,, \pi^{kl} (t^{\prime} ,
\underline{x}^{\prime} )\} &=& {{\delta}_{(ij)}}^{kl}\,
\alpha(t')\delta ( t , t^{\prime})\, \delta^3 ( \underline{x} ,
\underline{x}^{\prime}) \label{GR3b} \\
\{N(t,\underline{x}), p(t',\underline{x}')\} &=& \alpha(t)
\delta(t,t')
\delta^3(\underline{x}', \underline{x'}) \label{cansNp}\\
\{N(t,\underline{x}), N(t', \underline{x'}) \} &=& 0 \label{cansNN}\\
\{p(t,\underline{x}), p(t', \underline{x'}) \} &=& 0 \label{canspp} \\
 \{ N^i(t, \underline{x}), p_j(t',\underline{x}') \} &=& \delta^i_j
 \alpha(t)
\delta(t,t')\delta^3(\underline{x}', \underline{x'}) \label{canvNp} \\
\{N^i(t,\underline{x}), N^j(t', \underline{x'}) \} &=& 0
\label{canvNN} \\
\{p_i(t,\underline{x}), p_j(t', \underline{x'}) \} &=& 0 \,,
\label{canvpp}
\end{eqnarray}
where we have defined ${{\delta}_{(ij)}}^{kl} :=
\frac{1}{2}({\delta_i}^k {\delta_j}^l + {\delta_i}^l {\delta_j}^k
)$, and where $\alpha(t)$ has been defined earlier. All quantities
$N,N^i,p$ and $p_i$ have vanishing Poisson brackets with
$\pi^{ij}$ and $h_{ij}$.

\subsubsection{The Dirac algebra of constraints}
The construction above leads naturally to a one-parameter family
of Dirac super-hamiltonians $t\mapsto{\cal{H}}_{\bot} ( t,
\underline{x})$ and super-momenta $t\mapsto {\cal{H}}_i ( t,
\underline{x} )$. In the standard canonical approach to general
relativity\cite{ADM,Kuc91,I92}, the super-hamiltonian and
super-momenta are
\begin{eqnarray}
{\cal{H}}_{\bot} &=& \kappa^2 h^{-1/2}(\pi^{ij}\pi_{ij}
- \frac{1}{2} (\pi_i{}^i)^2) - \kappa^{-2}h^{1/2} R          \\
{\cal{H}}^i  &=& - 2 {\nabla}_{\!\!j} \pi^{ij}, \label{ADMHi}
\end{eqnarray}
where $\kappa^2=8\pi G/c^2$ and $\nabla$ denotes the  spatial
covariant derivative. We note that both these quantities are
spatial scalar densities, hence they can be smeared with scalar
quantities.

The history analogue of these expressions is
\begin{eqnarray}
{\cal H}_\perp(t,\underline x)&:=&\kappa^2 \alpha^{-1}(t)
h^{-1/2}(t,\underline{x})(\pi^{ij}(t,\underline{x})\pi_{ij}
(t,\underline{x})
- \frac{1}{2} (\pi_i{}^i)^2(t,\underline{x})) - \nonumber \\
&&\kappa^{-2}h^{1/2}\alpha(t)(t,\underline{x})
R(t,\underline{x}) \label{HperpHis}\\
\hspace{-1cm} {\cal{H}}^i(t,\underline x)&:=& - 2 {\nabla}_{\!\!j}
\pi^{ij}(t,\underline x).
\end{eqnarray}
We have introduced the weight $\alpha(t)$ in order to render the
determinant $h$ a density of weight zero with respect to time.

For each choice of the weight function $\alpha$, these quantities
on $\mathR\times\Sigma$ satisfy the history version of the Dirac
algebra
\begin{eqnarray}
\hspace{-1cm}\{ {\cal{H}}_i (t,\underline{x})\,, {\cal{H}}_j (
t^{\prime} , {\underline{x}} ^{\prime} )\}\ &=& - {\cal{H}}_j
(t,\underline{x}) \,\delta ( t , t^{\prime})\alpha(t')\,
{\partial^{x^{\prime}}}\!\!_{i} \,\delta^3 ( \underline{x}
,\underline{x}^{\prime}) \nonumber \\
 &&+ {\cal{H}}_i (t,\underline{x})\, \delta ( t , t^{\prime})\alpha(t')\,
{\partial^x}\!\!_{j}\, \delta^3 ( \underline{x} ,
\underline{x}^{\prime})  \label{Diracsm1}\\
\hspace{-1cm}\{ {\cal{H}}_i (t,\underline{x})\,, {\cal{H}}_{\bot}
( t^{\prime} , {\underline{x}}^{\prime} ) \} &=&  {\cal{H}}_{\bot}
(t,\underline{x})\, \delta ( t , t^{\prime})\alpha(t')\,
{\partial^{x^{\prime}}}\!\!_i \,\delta^3
( \underline{x} , \underline{x}^{\prime}) \label{Diracsm2} \\
\hspace{-1cm}\{ {\cal{H}}_{\bot} (t,\underline{x})\,,
{\cal{H}}_{\bot} ( {\underline{x}}^{\prime} , t^{\prime} )\} &=&
h^{ij}\! (t,\underline{x}) \,{\cal{H}}_i (t,\underline{x})\,\delta
( t , t^{\prime})\alpha(t')\, {\partial^{x^{\prime}}}\!\!_{j}\,
\delta^3 ( \underline{x} ,
\underline{x}^{\prime}) \nonumber  \\
 &&\!- h^{ij}\! ( t^{\prime} , {\underline{x}} ^{\prime} )\,{\cal{H}}_i
( t^{\prime} , {\underline{x}} ^{\prime} )\,\delta ( t ,
t^{\prime})\alpha(t')\,{\partial^x}\!\!_{j}\, \delta^3 (
\underline{x} , \underline{x}^{\prime}).  \label{Diracsm3}
\end{eqnarray}
Note, that when we introduce back the variables $\tilde{\pi}^{ij}$
and $\tilde{h}$ that are densities of weight 1 with respect to
time, the dependence on $\alpha(t)$ drops out from the expressions
for the constraints.

The smeared form of the super-hamiltonian ${\cal{H}}_{\bot}
(t,\underline{x})$ and the super-momentum ${\cal{H}}_i
(t,\underline{x})$ history quantities are defined using as their
smearing functions, respectively, a scalar function $L$, and a
spatial vector field $L^i$ as follows:
\begin{eqnarray}
{\cal{H}} (L) := \int d^3 \underline{x} \int dt\,\alpha(t)^{-1}
L(t,\underline{x})
{\cal{H}}_{\bot} (t,\underline{x})     \\
{\cal{H}} (\vec{L}) := \int d^3 \underline{x} \int
dt\,\alpha(t)^{-1} L^i (t,\underline{x}){\cal{H}}_i
(t,\underline{x}).
\end{eqnarray}

The smeared form of this history version of the Dirac algebra is
\begin{eqnarray}
\!\!\{ {\cal{H}}[ \vec{L} ]\,, {\cal{H}}[\vec{L^{\prime}}]\}\
&=& {\cal{H}} [ \vec{L} \,\,, \vec{L^{\prime}} ] \label{NiN'i} \\
\!\!\{ {\cal{H}} [ \vec{L} ] \,, {\cal{H}} [ L ] \} &=& {\cal{H}}
[{{\cal L}_{\vec{L}}} L ]  \label{H(NiN)} \\
\!\!\{ {\cal{H}}[ L ]\,, {\cal{H}} [L^{\prime}] \} &=& {\cal{H}} [
\vec{K}], \label{H(NN')}
\end{eqnarray}
where in Eq.\ (\ref{H(NN')}) we have $K^i := h^{ij}(L \partial_j
L' - L' \partial_j L)$, with $i=1,2,3$.

Hence, because the generators ${\cal H}_\perp(t,\underline x)$ and
${\cal{H}}^i(t,\underline x)$ of the history Dirac algebra Eqs.\
(\ref{Diracsm1}--\ref{Diracsm3}) trivially commute with the
variables $N$, $N^i$ $p$ and $p_i$ of the history algebra Eqs.\
(\ref{GR1b}--\ref{canvpp}), we recover exactly the history version
of the Dirac algebra. Therefore, on the history space $\Pi^{cov} =
\Pi^{can}$ we have a representation of the Dirac algebra {\em
together\/ with\/} a representation of the spacetime
diffeomorphisms group ${\rm Diff}(M)$.

This result is different from the one we obtained in an earlier
paper \cite{SavGR01}, in the sense that the 3+1 decomposition here
is obtained by means of the metric-dependent foliation. It was not
{\em a priori} evident that the results would stay the same. The
conclusion is that the structure of the canonical theory is not
affected by the introduction of the metric-dependent foliation,
but the latter is crucial if the canonical theory is to preserve
the spacetime character of the theory, namely the Lorentzian
nature of the spacetime metric.

\section{Conclusions}
In this paper, we have placed special emphasis on the dual nature
of the histories theory, that allows the comparison of spacelike
and canonical objects, as well as the explicit study of the
different choices of foliation.

We showed  that the histories theory  preserves the spacetime
character of the canonical description of general relativity. The
introduction of the metric-dependent foliation solved the problem
of the loss of the spacelike character of the foliation associated
to the 3+1 decomposition.  This allows the derivation of the exact
relation between the spacelike (covariant) and the canonical
descriptions of the histories general relativity theory.

We  concluded with the rather significant result that a
representation of both the group of spacetime diffeomorphisms and
the Dirac algebra of constraints co-exists in the metric-dependent
description of histories theory.

The detailed study of the symmetries of the theory, the
construction of the  reduced state space, and the histories
treatment of the problem of time,
 are studied in a continuation of this work
\cite{Sav03b}, which culminates in an explicit demonstration of
the ${\rm Diff}(M)$-invariance of canonical general relativity.

\vspace{1cm}

\noindent{\large\bf Acknowledgements}

\noindent I would like to especially thank Charis Anastopoulos for
very helpful discussions throughout this work. I would like to
also thank Chris Isham for his help in the editing of this paper.
I gratefully acknowledge support from the EPSRC GR/R36572 grant.

\vspace{0.7cm}

\begin{appendix}

\section{Foliations: some definitions}
A foliation of a four-manifold $M$ of topology $\Sigma \times
\mathR$ by three-surfaces $\Sigma$ is defined as a map
\begin{eqnarray}
 {\cal{E}} : {\Sigma \times \mathR } &\mapsto& M  \\
           (\underline{x} , t) & \rightarrow & {\cal{E}}(\underline{x}
           , t) := {{\cal{E}}_t} (\underline{x}).
\end{eqnarray}

Associated to such a foliation is an one-parameter family of
embeddings
\begin{eqnarray}
 {\cal{E}}_t : \Sigma &\mapsto& M   \\
             \underline{x} &\rightarrow&
             {\cal{E}}_t(\underline{x}),
\end{eqnarray}

The submanifolds $\Sigma_t$ of $M$ defined as $\Sigma_t =
{\cal{E}}_t(\Sigma)$, for each $t$ are known as the {\em leaves }
of the foliation {\cal{E}}. The choice of a foliation allows the
selection of a coordinate system common to all $\Sigma_t$
three-surfaces, in the sense that the coordinate defined on the
reference three-surface $\Sigma$ is shared by all $\Sigma_t$
three-surfaces.

For a coordinate system $x^i$ on $\Sigma$, where $i = 1,2,3$, the
three vector fields ${{\cal{E}}^{\mu}}_i$, tangent to the
foliation, are defined as
\begin{equation}
{{\cal{E}}^{\mu}}_i = \frac{\partial}{\partial x^i}{\cal{E}}^{\mu}
({\cal{E}}^{-1}(X)). \label{f5}
\end{equation}

Transverse to the leaves of the foliation is the deformation
vector, which is defined as
\begin{equation}
 t^{\mu}(X) = \frac{\partial}{\partial
 t}{\cal{E}}^{\mu}({\cal{E}}^{-1}(X)). \label{f4}.
\end{equation}

The vector fields ${{\cal{E}}^{\mu}}_i$ and $t^{\mu}$ form a
coordinate basis, so they satisfy
\begin{eqnarray}
 [ \;t, \;{{\cal{E}}_i} {\]^{\mu}} &=& 0    \\
 \[ {{\cal{E}}_i}, {{\cal{E}}_j} {\]^{\mu}} &=& 0.
\end{eqnarray}

\end{appendix}

\end{document}